\documentclass[12pt,a4paper]{article}

\usepackage{epsfig}

\newcommand{\bi}{\begin{itemize}}
\newcommand{\ei}{\end{itemize}}

\newcommand{\benu}{\begin{enumerate}}
\newcommand{\eenu}{\end{enumerate}}

\newcommand{\be}{\begin{equation}}
\newcommand{\ee}{\end{equation}}

\newcommand{\bea}{\begin{eqnarray}}
\newcommand{\eea}{\end{eqnarray}}

\def\lsim{\ \raisebox{-.45ex}{\rlap{$\sim$}} \raisebox{.45ex}{$<$}\ }
\def\gsim{\ \raisebox{-.45ex}{\rlap{$\sim$}} \raisebox{.45ex}{$>$}\ }

\newcommand{\THK}{\mbox{\sf T2K}}
\newcommand{\NOVA}{\mbox{\sf NO$\nu$A}}
\newcommand{\HK}{\mbox{\sf HK}}
\newcommand{\SK}{{\sf SK}}
\newcommand{\UNO}{\mbox{\sf UNO}}

\newcommand{\ie}{{\it i.e.}}

\newcommand{\eq}{Eq.}

\newcommand{\fig}{Figure~}

\newcommand{\Tab}{Table~}

\def\numu{{\nu_{\mu}}}
\def\anumu{{\bar\nu_{\mu}}}

\def\evsq{{{\rm eV$^2$}}}

\def\da{{\Delta_{31}}}
\def\ds{{\Delta_{21}}}

\def\dcp{{\delta_{\mathrm{CP}}}}

\def\stsmall{{\sin^2 2 \theta_{13}}}

\def\pmutau{{{\rm P_{\mu \tau}}}}
\def\pmue{{{\rm P_{\mu e}}}}
\def\pmumu{{{\rm P_{\mu \mu}}}}
\def\boldpmumu{{{\rm {\bf{P_{\mu \mu}}}}}}

\def\pamumu{{{\rm P_{\bar \mu \bar \mu}}}}

\def\numutonue{{{\rm {\nu_\mu \to \nu_e}}}}

\def\sumnumu{{\nu_{\mu} + \bar\nu_{\mu}}}

\def\signda{{{{\sf{sign}}}}{(\Delta_{31})}}

\newcommand{\chr}{\mbox{$\breve{\rm C}$erenkov~}}
\newcommand{\boldchr}{\mbox{$\breve{\rm {\bf C}}$erenkov~}}

\begin{document}


\begin{flushright}
{\makebox[1.cm] 
        \sf hep$\,$-$\,$ph/0506145}
\end{flushright}

\vspace*{1cm}

\renewcommand{\thefootnote}{\fnsymbol{footnote}}

\setcounter{footnote}{-1}

{\begin{center}
{\Large\textbf{
Probing the $\nu$ Mass Hierarchy
via Atmospheric $\nu_\mu + \bar \nu_\mu$ Survival Rates in Megaton
Water \boldchr Detectors
}}
\end{center}}

\vspace*{.8cm}

{\begin{center} {{\bf
                Raj Gandhi $^{a, \,\!\!\!}$
\footnote[1]{\makebox[1.cm]{Email:}
                \sf raj@mri.ernet.in},
                Pomita Ghoshal $^{a, \,\!\!\!}$
\footnote[2]{\makebox[1.cm]{Email:}
                \sf pomita@mri.ernet.in}, 
                Srubabati Goswami $^{a, \,\!\!\!}$
\footnote[3]{\makebox[1.cm]{Email:}
                \sf sruba@mri.ernet.in},
           Poonam Mehta $^{a,b,}$
\footnote[4]{\makebox[1.cm]{Email:}
                \sf mpoonam@mri.ernet.in, 
                    poonam@wisemail.weizmann.ac.il},
   S Uma Sankar $^{c, \,\!\!}$
\footnote[5]{\makebox[1.cm]{Email:}
                \sf uma@phy.iitb.ac.in}
                }}
\end{center}}
{\it
\begin{center}
       $^{a}$Harish-Chandra Research Institute,
 Chhatnag Road, Jhunsi,\\
Allahabad 211 019, India\\[4mm]
       $^{b}$Department of Particle Physics, 
Weizmann Institute of Science,\\
Rehovot 76 100, Israel
\\[4mm]
       $^c$Department of Physics, Indian Institute of Technology, 
Powai,\\
Mumbai 400 076, India
\end{center}}

\vspace*{0.5cm}

\date{today}


\begin{abstract}

The neutrino mass hierarchy, presently unknown, is a 
powerful discriminator among various classes of unification 
theories. We show that the $\sumnumu$ survival rate in
atmospheric events can provide a novel method 
of determining the hierarchy in megaton water \chr detectors. For
pathlength and energy ranges relevant to atmospheric neutrinos, 
this rate obtains significant matter sensitive variations not 
only from resonant matter effects in $\pmue$ but also from 
those in $\pmutau$. 
We calculate the expected muon event rates in the case of matter 
oscillations with both natural and inverted hierarchy. We 
identify the energy and pathlength ranges for which resonant matter
effects can lead to observable differences between the above two
cases. We also estimate the exposure time required to observe
this difference and  determine the sign of $\Delta_{31}$
in a statistically significant manner. 
%
\end{abstract}


\newpage

\renewcommand{\thefootnote}{\arabic{footnote}}

%
%

\setcounter{footnote}{0}


\section{Introduction}

An oft-repeated statement is that the recent evidence for neutrino
mass and oscillations \cite{sru} constitutes proof of physics beyond
the Standard Model. The important sub-text to this, however, is that
the nature of this physics remains, at present, largely unknown, with very many
possibilities and open questions. The quest for this new physics  is
closely linked to several unknowns in the neutrino sector :   
\benu
\item[{\sf {(a)}}] The absolute scale of neutrino masses,
\item[{\sf {(b)}}] The hierarchy of neutrino masses, 
\item[{\sf {(c)}}] The magnitude of the lepton mixing matrix  
element ${U_{e3}}$ (see below), 
\item[{\sf {(d)}}] The value of the leptonic CP violating phase $\dcp$, and 
\item[{\sf {(e)}}] The Dirac or Majorana nature of neutrinos.
\eenu

The {\it {hierarchy of neutrino masses }} $m_i$, ((b) above) with $i=1,2,3$, refers to the
ordering of the mass eigenstates. Given the fact that
$\ds = m^2_2 - m^2_1 > 0$ from solar neutrino data, knowing 
the hierarchy translates to determining the (as yet unknown\footnote{
Our knowledge of $\da = m^2_3 - m^2_1 $ 
derives from atmospheric and long baseline accelerator (K2K) 
data, which are sensitive only to its magnitude.})
sign of $\da = m^2_3 - m^2_1$, henceforth referred to as $\signda$.
Specifically, if this sign is positive, 
({\it i.e.} $m_3^2 > m_2^2 > m_1^2$)
the hierarchy is termed {\it normal} (NH), 
and if it is negative ({\it i.e.} 
$m_2^2 > m_1^2 > m_3^2$), 
the hierarchy is said to be {\it inverted} (IH).
The neutrino mixing matrix 
$U$, with elements $U_{\alpha i}$
 relating the flavour or weak interaction eigenstates (labeled by
$\alpha = e,\mu,\tau$) to the mass eigenstates (labeled by $i=1,2,3$) 
leads to the same flavour composition for a given mass state  whether the 
hierarchy is normal or inverted. 
Attempts to construct a unified theory beyond the Standard Model however, 
depend crucially on $\signda$; in fact,
one way to classify families of models is via 
the neutrino hierarchy they assume as input.

A large class of Grand Unified Theories (GUTs) use the 
Type I seesaw mechanism \cite{moh}, requiring the existence of 
heavy right-handed neutrinos at the GUT scale to generate small 
neutrino masses. Since GUTs typically unify quarks and leptons, 
and the quark hierarchy is normal, such models favour a normal 
neutrino hierarchy. 
Typically, if one uses an inverted hierarchy to construct a 
Type I seesaw model, one finds \cite{alb} that the light neutrino 
mass spectrum may become very sensitive to small changes in the
heavy neutrino masses and to radiative corrections. 
Inverted hierarchies imply the near degeneracy of the 
states $m_2$ and $m_1$, and since a corresponding degeneracy  is
absent in the quark sector, they may  require the presence 
of an additional global symmetry in the lepton sector.
Additionally, important features of leptogenesis,
which may help partially understand the present
matter-antimatter  asymmetry of the universe, are 
significantly different in
the two cases. In general, GUTs based on Type I seesaw models 
strongly depend on a normal mass hierarchy to 
obtain their many desirable features\cite{alb}.
An inverted hierarchy, on the other hand, would favour the
possibility of a unified theory based on a Type II seesaw
mechanism, employing additional Higgs triplets, or on models which
require a (global) lepton flavour symmetry.
The hierarchy is thus a crucial marker in our quest
for a unified theory, and its determination would 
be helpful in eliminating 
or at the very least strongly disfavouring large classes of such theories and 
in considerably narrowing the focus of this search.

Generally speaking, determination of the mass hierarchy requires
the observation of resonant matter effects ({\it i.e.} long
baselines) and a not too small $\stsmall ( \gsim ~0.05 )$\footnote{While these conditions are sufficient, they are
not necessary. If they  are not satisfied, however, qualitatively
different experimental approaches entailing a significantly higher
degree of precision will be required \cite{andre}.}. 
Among the next generation long baseline accelerator experiments, combined
results from \THK ~\cite{t2k,hk} and the NuMI Off-axis 
experiment \NOVA ~\cite{numi} may be able to infer the neutrino mass hierarchy
\cite{manfred10,Minakata:2003ca,Mena:2004sa, Mena:2005sa}. It is 
also possible to determine the hierarchy in experiments 
using beta beams \cite{Burguet-Castell:2003vv} and 
neutrino factories \cite{Barger:2000cp}.
 However these experiments use the $\nu_\mu \leftrightarrow \nu_e$ 
channel, the sensitivity of which is compromised by  the
ambiguities resulting from inherent degeneracies. Specifically,
these originate in inter-relations between the $\signda$, 
the phase $\dcp$ and 
the mixing angle $\theta_{13}$ \cite{degeneracy}.
 To overcome this, the synergistic use of two experiments, 
sometimes with more than one measurement each will be necessary
\cite{Barger:2002xk,Burguet-Castell:2002qx,
Wang:2001ys,Whisnant:2002fx,Huber:2003pm,Donini:2004hu}.
Alternatively, two detectors at different distances
\cite{ishitsuka,mena} have been suggested. For baselines
contemplated for the next generation of accelerator and superbeam
experiments (\ie $\lsim 800$ km), much of the decrease in
sensitivity arises from the $\dcp$ - $\signda$
degeneracy. Its effect tends to decrease for much longer
baselines, vanishing, in fact, for  the magic baseline 
where the $\dcp$ terms go to zero \cite{Huber:2003ak,
Asratyan:2003dp, arc}.
The  use of earth matter effects in  atmospheric muon
neutrinos for the determination of the hierarchy has been
studied in the context of magnetized iron calorimeter 
detectors in \cite{palom,indu,us2} and water \chr 
detectors in \cite{mue-h}. 
In particular, it was shown in \cite{us2}
that the effect of degeneracies are not significant in the muon
survival probability for the large baselines involved.
Matter effects in supernova neutrinos can
also, in principle, determine  the neutrino mass hierarchy
\cite{smir}. 
However large uncertainties in supernova neutrino
fluxes reduce the sensitivity. 
Finally, if neutrinos are Majorana particles then determination 
of mass hierarchy may be possible from next generation neutrino-less 
double beta experiments \cite{pascoli}, provided hadronic 
uncertainties in the nuclear matrix elements are reduced from 
their present levels. 
Non-oscillation probes of the neutrino mass hierarchy and the 
limit of small $\stsmall$ have been discussed recently in \cite{andre1}.

Megaton water \chr detectors constitute an important
future class of detectors with significant capabilities. 
The projects under active consideration are \UNO 
~\cite{uno} in the US, a megaton detector in the Frejus
laboratory \cite{frejus} in Europe, and the 
HyperKamiokande (\HK) project \cite{hk,kaj} in Japan. 
Such massive detectors would enable impressive 
measurements of atmospheric neutrino parameters. 
Specifically, they would determine $|\da|$ and $\sin^2 \theta_{23}$ 
to the few percent level, and obtain a much improved bound on $\theta_{13}$.
Since the water \chr technique is not sensitive to the charge 
of the produced lepton on an event by event basis, the combined
signal for neutrinos and antineutrinos must be searched for using statistical
discriminators in order to determine the mass hierarchy. 
At present, the method which is contemplated for discriminating 
between neutrino and antineutrino interactions involves using the 
differences in the total cross section $\sigma$ and its 
rapidity\footnote{Rapidity is defined as 
${\mathrm{y = (E_{\nu} - E_{lepton})/E_{\nu}}}$.} dependence ($d\sigma/dy$). 
Neutrino-nucleon interactions in the few GeV to 10 GeV range have a 
higher average rapidity than the corresponding antineutrino-nucleon 
interactions, and thus produce more multi-hadron events. 
Thus an enhanced signal for  multi-ring electron-like events is expected 
in water \chr detectors if the hierarchy is normal than if it is 
inverted \cite{mue-h,kaj}. 
Given the importance of the  hierarchy to theoretical efforts towards
unification, it is recommended that we explore multiple approaches 
to this problem. 
In this paper, we exploit the sensitivity of the
muon survival rate to  {\it resonant} matter effects (as opposed to merely
matter {\it enhanced} effects) for detecting the 
hierarchy in megaton \chr detectors. 
Our method consists of a careful selection of energy and 
zenith angle bins for which the resonant matter effect in muon neutrino 
survival probability is observationally significant. 
We perform realistic event-rate calculations for atmospheric muon 
rates in such detectors, which incorporate their efficiencies and 
lie within their resolutions. 
We show that for $\stsmall = 0.1$ and $\sim$ a 3 year 
run, signals with statistical significance in excess of 
$4 \sigma$ are possible
for pathlength and energy ranges where the muon survival
probability is matter sensitive. 
As mentioned above, the results are largely free of parameter degeneracies 
and associated ambiguities.


\section{Earth - matter effects in $\boldpmumu$}

As is well known the matter effects in neutrino oscillations arise
due to the difference in the interactions of $\nu_\mu$/$\nu_\tau$ 
and $\nu_e$ as they traverse matter \cite{wolf}. The mixing angle
$\theta_{13}$ parameterizes  the admixture of $\nu_e$ with 
$\nu_\mu$/$\nu_\tau$ in the oscillations driven by the larger 
mass-square difference $\da$. Hence all matter effects in  
oscillations, such as those which manifest themselves in long baseline  and 
atmospheric neutrino experiments, are proportional to sine of this angle. 
The CHOOZ experiment constrains  $\sin^2 2 \theta_{13}$ to be $ \leq 0.2$ \cite{chooz}. For the 
energy which satisfies the resonance condition  \cite{mikhsmir}
\begin{equation}
\da \cos 2 \theta_{13} = 0.76 \times 10^{-4} \mathrm{\rho(gm/cc)~E(GeV)}
\label{rescon}
\end{equation}
the matter dependent $\theta_{13}$ gets amplified to $\pi/4$ and
matter effects lead to large changes in the survival/oscillation
probabilities, provided the pathlength dependent oscillating term
is also close to $1$. 

Resonant matter effects involving 
$\numutonue$ transitions
have been extensively studied in the literature \cite{numue}. 
Most studies of next generation accelerator neutrino experiments,
seeking to probe the mass hierarchy seek to exploit the matter effects
in $\pmue$. 
These experiments have baselines ranging from 300 km 
to 3000 km and energies in the $\sim$ GeV range. 
For such `short' baselines and for energies in the few GeV range, 
while the percentage change in $\pmue$ can be large, its absolute value remains modestly small,
since {\it{resonant}} matter effects do not develop. 
However, for the baseline range 6000 - 10500 km, since $\rm{\sin^2 2\theta^m_{13}}$
then gradually builds to $\sim$ 1, $\pmumu$ and $\pmutau$ exhibit
large matter sensitive variations, expecially if resonance occurs
in the neighbourhood of a vacuum peak. 
In such a situation, the oscillating factors in the 
survival/oscillation probabilities are large. This in turn synergistically converts
a large change in $\stsmall$ due to matter effects into a large change in survival/oscillation
probabilities.
 
It is difficult to observe the {\it resonant} amplification of the 
matter effects in $\numutonue$ oscillations, because the 
pathlength for which this  occurs is inversely 
proportional to $\tan 2 \theta_{13}$.
For almost all allowed values of $\theta_{13}$, except those near the 
current upper bound, this pathlength is close to or  larger than the diameter of
earth \cite{us1}:
\begin{equation}
\mathrm{[\rho L]_{\mu e}^{max}} \simeq  \frac{(2p+1) \pi   
\times 10^4}{2\tan 2 \theta_{13}} \ {\mathrm{km~gm/cc}},
\label{Lmue}
\end{equation} 
where $\rho$ is the average density of the earth along the
path and $p$ is a positive integer $\geq$ 0.
However, the pathlength 
for which resonant matter effects in $\pmumu$ are maximum 
is given by \cite{us1}
\begin{equation}
\mathrm{[\rho L]_{\mu\mu}^{max}} \simeq p \pi \cos 2 \theta_{13} 
\times 10^4 \ {\mathrm{km~gm/cc}}, 
\label{defL} 
\end{equation}
where $p$ is a positive integer $>$ 0.
For all allowed values of $\theta_{13}$, $\cos 2
\theta_{13}$ is close to 1.
Thus the resonant amplification in $\pmumu$ occurs within a narrow range of pathlengths for all the allowed values of $\theta_{13}$,
thus making $\pmumu$ a suitable observable to study it.
We obtain this pathlength to be about 7000 km by 
substituting the mantle density of earth $\rho = 4.5$ gm/cc  
and $p = 1$ in Eq.(\ref{defL}). From a numerical study we have
identified the energy range 5 - 10 GeV and the pathlength 
range 6000 - 9000 km to be suitable for studying resonant amplification
of matter effects in $\pmumu$. In the case of $\pmutau$,
the pathlength
for which
matter effects cause the largest change is given by \cite{us1}
{\footnote{Note that in \cite{us1} $\cos 2\theta_{13}$ 
is misprinted as $\cos^2 \theta_{13}$. hep-ph/0408361, v3 has the correct expression.}} 
\begin{equation}
\mathrm{[\rho L]_{\mu\tau}^{max}} \simeq \frac{(2p+1)}{2} \pi 
\cos 2 \theta_{13} 
\times 10^4 \ {\mathrm{km~gm/cc}}. 
\label{Lmutau}
\end{equation}
Here again the resonance amplification occurs for essentially
the same pathlength for all allowed values of $\theta_{13}$.
Setting $p = 1$, in Eq.(\ref{Lmutau}), we get this pathlength
to be about 9700 km{\footnote{For $p=0$, or the $\pi/2$ peak, the matter-induced change in $\pmutau$
is seen to be small \cite{us2}.}}. 
A numerical study shows that the energy
range 4 - 8 GeV and pathlength range 8000 - 10500 GeV is suitable
to study matter effects in $\pmumu$ induced by large matter-induced changes in $\pmutau$.

\begin{figure*}[!htb]
\vskip 1cm
{\centerline{
\hspace*{2em}
\epsfxsize=12cm\epsfysize=9cm
                     \epsfbox{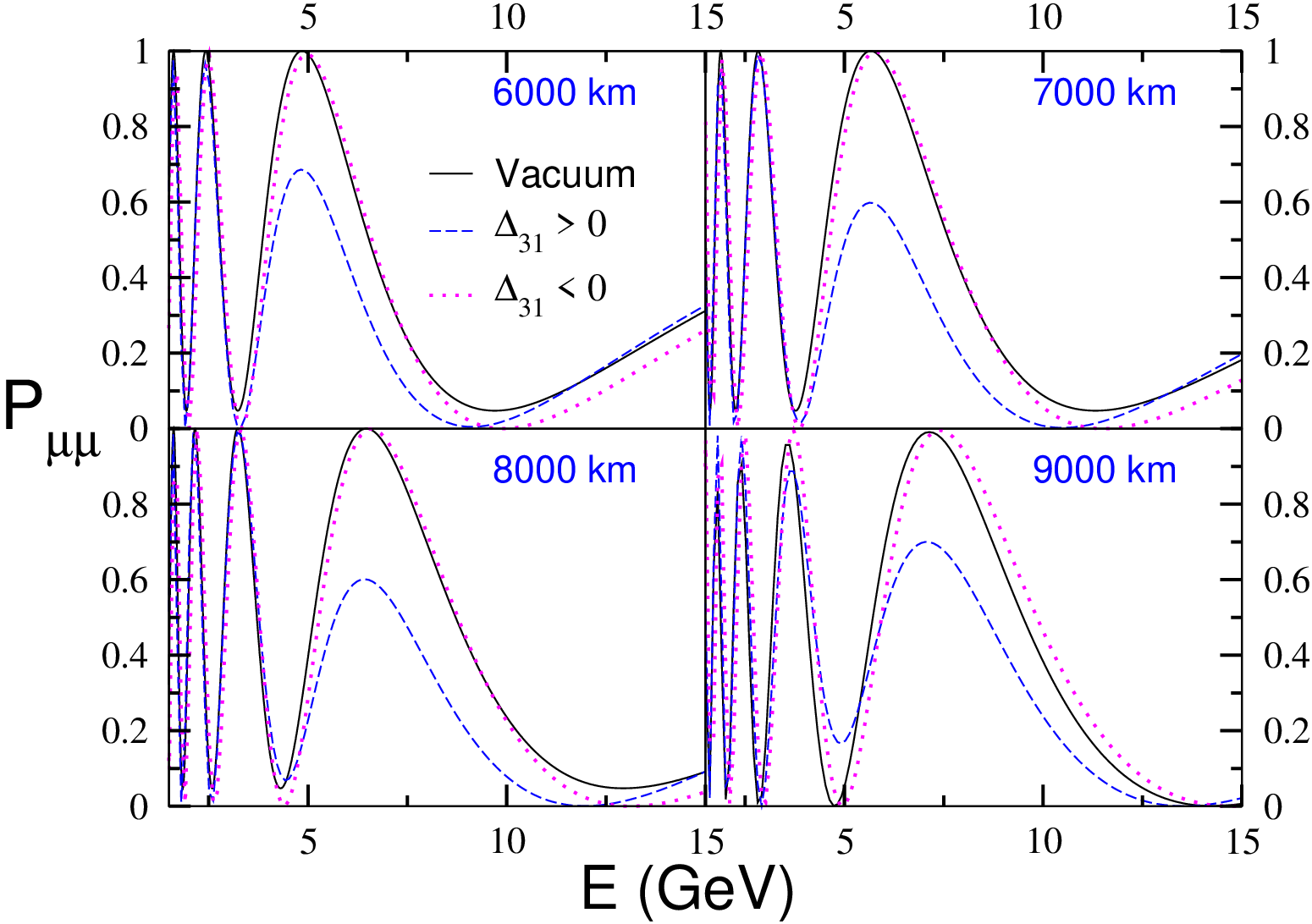}
}
\caption{\footnotesize
The muon survival probability $\pmumu$ in matter 
plotted as a function of E (GeV) for four pathlengths,
L = 6000, 7000, 8000 and 9000 
km and  
for the two signs of $\da$. 
The values of parameters used are 
$\sin^2 \theta_{12} = 0.28$, $\sin^2 \theta_{23} = 0.5$, 
$\stsmall = 0.1$, $\rm{\vert \da \vert = 2 \times 10^{-3}}$ 
\evsq~ and $\rm{\ds = 8.3 \times 10^{-5}}$ \evsq. 
The survival probability in 
vacuum is also shown for comparison.
}
\label{fig1}
}
\end{figure*}

\begin{figure*}[!htb]
\vskip 1cm
{\centerline{
\hspace*{2em}
\epsfxsize=12cm\epsfysize=9cm
                     \epsfbox{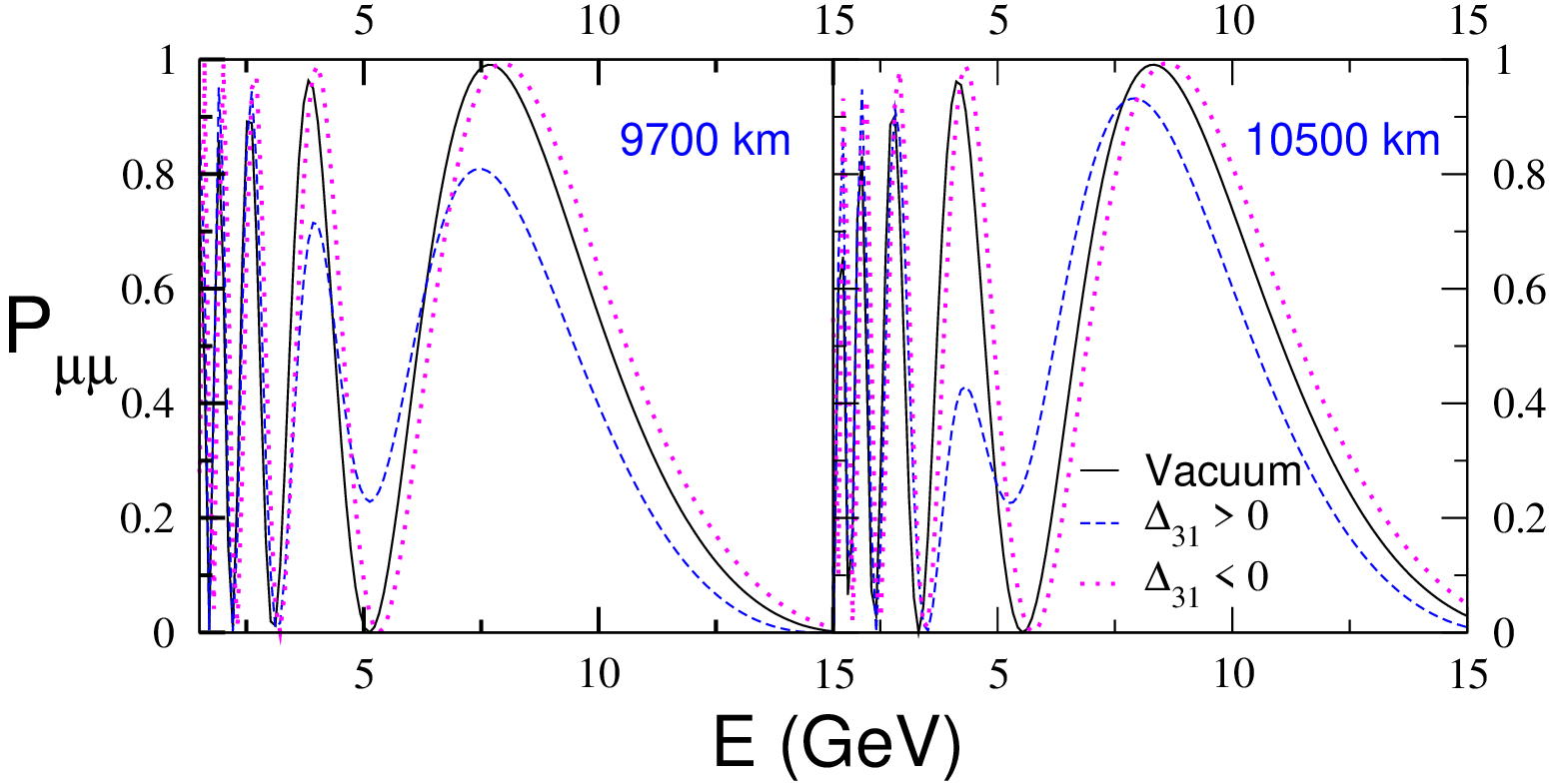}
}
\caption{\footnotesize
The muon survival probability $\pmumu$ in matter 
plotted as a function of E (GeV) for two pathlengths, 
L = 9700 and 10500 km and 
for the two signs of $\da$. 
The values of parameters used are 
$\sin^2 \theta_{12} = 0.28$, $\sin^2 \theta_{23} = 0.5$, 
$\stsmall = 0.1$, $\rm{\vert \da \vert = 2 \times 10^{-3}}$ 
\evsq~ and $\rm{\ds = 8.3 \times 10^{-5}}$ \evsq. 
The survival probability in 
vacuum is also shown for comparison.
}
\label{fig2}
}
\end{figure*}

For the long pathlengths under consideration here, 
we need to explicitly take into account the 
varying density profile of the earth. 
We use the density profile given in Preliminary Reference 
Earth Model (PREM) \cite{prem}. 
In \fig\ref{fig1} we plot $\pmumu$ as a function of energy for
four different pathlengths in the range 6000 to 9000 km. In all of
these four cases, the most significant matter effects occur in 
the energy range 5 - 10 GeV. 
In \fig\ref{fig2} we plot $\pmumu$ as a function of energy for
the two pathlengths 9700 km and 10500 km. In both these cases 
the most significant matter effects occur in the energy range
4 - 6 GeV. 
The curves \fig\ref{fig1} and \fig\ref{fig2} have been obtained 
by numerically solving the full three flavour
neutrino propagation equation through earth matter. In obtaining
these curves we have used the following values for neutrino parameters:
$\vert \da \vert = 0.002$ \evsq, $\ds = 8.3 \times 10^{-5}$ 
\evsq, $\sin^2 \theta_{12} = 0.28$, $\sin^2 \theta_{23} = 0.5$
and $\sin^2 2 \theta_{13} = 0.1$.  
For each of the six pathlengths, the muon neutrino survival 
probability $\pmumu$ is 
calculated in vacuum and in matter for both the signs of $\da$. 


The following comments are in order :
\begin{itemize}
\item
For negative $\da$ (or inverted hierarchy), 
there is no discernible difference between
vacuum and matter survival probabilities.
\item
For positive $\da$ (or normal hierarchy), 
the value of $\pmumu$ in matter shown in \fig\ref{fig1} suffers a
drop with respect to its vacuum value in the energy range 5 - 10 GeV.
Whereas,   
the $\pmumu$ shown in \fig\ref{fig2} undergoes an
increase in the energy range 4 - 7 GeV.
\item
For the pathlength range 6000 - 9000 km, the change 
in $\pmumu$, due to matter effects, is dominated by 
change in $\pmue$.
In the vicinity of the vacuum peak, matter $\pmumu$ is
smaller by about $40 \%$ compared to the vacuum value.
Three fourths of this change occurs due to change in
$\pmue$ and the rest is due to change
in $\pmutau$ \cite{us2}. This is illustrated in 
\fig\ref{fig1}.
\item
For pathlengths $\gsim$ 9000 km the matter effect in $\pmutau$ also
becomes significant. For such pathlengths, there is a
drop in $\pmutau$ which is as high as 70\%, in the energy 
range 4 - 6 GeV. This drop in $\pmutau$ overcomes the rise
in $\pmue$. Thus the net change in $\pmumu$ is an increase of 
the matter value over its vacuum value \cite{us1,us2}. 
This is illustrated in \fig\ref{fig2} for two pathlengths
9700 km and 10500 km.
\item Beyond 10500 km the neutrinos start
traversing  the core and the mantle-core interference
effects set in \cite{petcov_nlo}. It was shown in \cite{us2} that
at such pathlengths and energies relevant for  atmospheric neutrinos
the difference between the vacuum and matter event rates are not
significant\footnote{In the energy range $\sim$ 3-6 GeV there is
some difference between matter and vacuum rates due to mantle-core
interference effects \cite{petcov-private}.}. Hence we do not
include such pathlengths in our analysis. 
\item In the  case of muon antineutrinos,  $\pamumu$ is essentially
unchanged for the case of positive $\da$. It will experience 
resonance enhanced suppression for the case of negative $\da$,
which again can be as large as $40 \%$ for pathlengths in the
range 6000 - 9000 km.  
\end{itemize}

It is clear from \fig\ref{fig1} and \fig\ref{fig2} that matter effects 
in $\pmumu$ can be large, and a high statistics  measurement of
the muon survival rate in the energy and pathlength ranges indicated can
be used to detect their presence as well as pin down the mass
hierarchy. Atmospheric neutrinos passing through earth's mantle
have the relevant pathlengths and possess energies in the 
desired range and thus are well-suited for this purpose.

In general, these effects are sensitive to $\sin^2 \theta_{13}$, as
mentioned earlier. This sensitivity, along with other aspects, was
recently studied for a charge discriminating detector 
in \cite{palom,indu,us2}. 
In \cite{us2}, it was emphasized that a conclusive and 
statistically significant determination of the hierarchy 
and associated matter effects in such a detector requires a careful 
selection of pathlength and energy range.
It was shown that a 4$\sigma$ signal for matter effects
is possible with an exposure of 1000 Ktyr in a
typical iron calorimeter detector \cite{monolith,ino}
when muon events in the energy range 5 - 10 GeV and the
pathlength range 6000 - 9700 km are considered.
Wider ranges of energies and pathlengths
result in an averaging out of the signal for 
matter effects \cite{us2}.
In what follows, we study the sensitivity of the
$\sumnumu$
survival rate to the hierarchy and to matter
effects for a megaton size water \chr detector.

\section{Numerical results for water \boldchr detectors}

In order to determine the type of neutrino mass hierarchy,
we exploit the muon neutrino and antineutrino
disappearance channels. 
For a normal hierarchy (NH), the resonant amplification of matter
effects occurs only for neutrinos, while for an inverted
hierarchy (IH) it occurs only for anti-neutrinos. 
From \fig\ref{fig1} and \fig\ref{fig2} we see that, for a  NH,
the $\mu^-$ rate undergoes a change due to matter effects
but the $\mu^+$ rate is essentially the same as the vacuum 
oscillation rate. In the case of an IH, the situation is reversed.
Here we have to make a distinction between two different  
kinematic choices:
\begin{enumerate}
\item
\underline{E = 5 - 10 GeV and L = 6000 - 9000 km}:
In this energy and pathlength ranges the decrease in
$\pmumu$ is induced mainly by the increase in $\pmue$.
For a NH, matter effects suppress the $\mu^-$ event 
rates by a large fraction (about $25 \%$) compared to their
vacuum oscillation rates whereas the $\mu^+$ event rates will
be essentially the same as their vacuum oscillation values.
For an IH, the $\mu^-$ rates are equal to their vacuum rates and 
the $\mu^+$ rates undergo significant suppression.
These results follow  from the plots of $\pmumu$
shown in \fig\ref{fig1}.
\item
\underline{E = 4 - 6 GeV and L = 9000 - 10500 km}:
In this energy and pathlength ranges $\pmumu$ increases
due to a sharp fall in $\pmutau$. 
For a NH, matter effects increase the $\mu^-$ rates compared
to their vacuum oscillation rates whereas, once again,
the $\mu^+$ rates are unaffected. For an IH, the situation
is reversed. Once again, these results are apparent in
 the plots shown in \fig\ref{fig2}.
\end{enumerate}
 
In a charge discriminating detector, which detects $\mu^-$
and $\mu^+$ rates individually, a comparison of the $\mu^-$ event
rate with that expected from vacuum oscillations is enough to
determine  $\signda$, whether the signal is measured 
in the ranges 1 or in ranges 2 \cite{us2}. 
Water \chr detectors are insensitive to  the sign of the leptonic charge  on
an event-by-event basis.
Therefore, we sum over the rates of $\mu^-$ and $\mu^+$
events and label the sum as `muon events'. Hence, for both
normal and inverted hierarchies, the total muon event rate due to matter 
oscillations will be less(more) than the corresponding rate in 
the case of vacuum oscillations in the case of ranges 1(2).

While the summation of events over muons and anti-muons dilutes  
the sensitivity of  water \chr detectors to matter
effects  compared to charge discriminating detectors, the proposed megaton mass
overrides this disadvantage and provides the statistics necessary for
a determination of the hierarchy. It is also to be noted that 
 the $\mu^-$ event
rates are $2$ to $3$ times larger than those of the  $\mu^+$ rates
due to the larger $\nu_\mu-N$ cross section. Thus
a larger change in the  total muon event rate and a statistically stronger signal of 
matter resonance effects is envisioned in the case of a
NH than in the case of an IH in both types of detectors.

The total number of muon
(or anti-muon) charged current (CC) events  can be obtained by  folding the
relevant  CC cross section with the survival probability,
the incident neutrino flux ,the  efficiency for detection, the detector mass
and the  exposure time.
In our calculations, the cross sections for $\rm{\numu-N}$ and 
$\rm{\anumu-N}$ interactions have been 
taken from \cite{sk2005}.
The total CC cross section is sum of quasi-elastic, 
single meson production and deep inelastic cross sections.

Before describing our calculation and the results, we first
state the assumptions we have made. We assume that  the 
neutrino oscillation parameters  $\vert \da \vert$ and $\theta_{23}$ will be 
determined to better than $10 \%$ accuracy by the long baseline
experiments MINOS \cite{min} and T2K \cite{t2k} by the time megaton size water \chr detectors are
operative. In addition, we have assumed that similar precision will exist for 
the measured values of $\Delta_{21}$ and $\theta_{12}$ \cite{sru,solar}, even though the 
dependence of our results on them is marginal. Based as they are on survival
rates and large matter effects, our conclusions are also effectively
independent of the 
 CP violating phase $\delta_{CP}$. Finally,  for our method to yield results 
over running times of 3-4 years, $\stsmall$ must be $\geq 0.05$.

For the first set of our calculations, we use the following values
for all neutrino parameters known to be non-zero: $\vert \da \vert =
0.002$ \evsq, $\Delta_{21} = 8.3 \times 10^{-5}$ \evsq,
$\sin^2 \theta_{23} = 0.5$ and $\sin^2 \theta_{12} = 0.28$.
We calculate the event rates for four different values of
$\theta_{13}$, i.e. $\sin^2 2 \theta_{13} = 0.05, 0.1, 0.15, 0.2$.
In a later set of calculations, we vary $\theta_{23}$ and $\vert \da \vert$ within the
ranges allowed by SK atmospheric neutrino data and K2K results to gauge the
dependence of our results on them.

To study the sensitivity of the megaton water \chr detectors
to the hierarchy, we compute the expected number of $\mu^- 
+ \mu^+$ events in the following four cases: 
\begin{itemize}
\item No oscillations,  $\mathrm{N_{Down}}$,
\item Vacuum oscillations, $\mathrm{N_{vac}}$,
\item Matter oscillations with NH,  $\mathrm{N_{NH}}$ and
\item Matter oscillations with IH, $\mathrm{N_{IH}}$.
\end{itemize}
If the difference between $\mathrm{N_{NH}}$ and 
$\mathrm{N_{IH}}$ is 
statistically significant, then $\signda$ can be assumed to
established with this same significance. Thus 
\begin{equation}
\mathrm{N_\sigma} = \frac{|\mathrm{N_{NH}} - \mathrm{N_{IH}}|
           }{\sqrt{\mathrm{N_{NH}}}} 
\label{Nsigma}
\end{equation}
gives the {\it sigma confidence level with which
the expectation for NH differs from that for IH}. 

We note that  Eq.(\ref{Nsigma}) tacitly  assumes that atmospheric neutrino
fluxes will be measured well enough in the next decade  to allow use of the
absolute event rates. However, if this is not the case,
 one can also
use the {\it measured} number of downward going events, which have 
the same corresponding values of $|\cos \theta_{\rm zenith}|$
as the upgoing events, as an estimate for the number of events 
in case of no oscillations. This  is the reason why we have denoted these as
$\mathrm{N_{Down}}$ above. In such
a situation, we can usefully consider the ratios $\mathrm{R_{NH}} = 
\mathrm{N_{NH}}/\mathrm{N_{Down}}$ and $\mathrm{R_{IH}} = 
\mathrm{N_{IH}}/\mathrm{N_{Down}}$. The statistical  
significance of the difference between the ratios is then defined as
\begin{equation}
\mathrm{N_{\sigma_R}} = \frac{|\mathrm{R_{NH}} - 
\mathrm{R_{IH}}|
        }{\mathrm{R_{NH}}\sqrt{1/\mathrm{N_{NH}} + 
1/\mathrm{N_{Down}}}}.
\label{Nsigma2}
\end{equation}

Our first set of results is for  SuperKamiokande (\SK) for 15 years 
of running. Since the \SK~ fiducial volume is 22.5 Kt, 
this gives us an exposure\footnote{
In the latest \SK~ paper \cite{sk2005}, the exposure
is 92 Ktyr
corresponding to 1489 days live time from May 1996 to July 2001.
Preliminary results for SK-II include a further 627 days live time \cite{suzuki}.
} 
of 337.5 Ktyr.
We use the Bartol fluxes from \cite{flux} and
incorporate the \SK~ cross sections, resolutions \cite{sk2005,sk2} and 
efficiencies \cite{kajita-private1}. In calculating the number of
events, we have integrated over the neutrino energy range 
= 5 - 10 GeV and pathlength range L = 6000 - 9000 km.
For $\da = 0.002$ \evsq, the resonance in earth's mantle occurs for
energy of about $6.5$ GeV. 
Since we are looking for resonant amplification of matter effects, 
we have  chosen the energy range so that it encompasses this
energy. We note that this choice is also dependant on the fact that
 $\vert \da \vert = 0.002$ \evsq. For other values of $\vert \da
\vert$, the appropriate energy range will be different. It was shown
earlier that the pathlength for which the resonant matter effects are
maximum is about 7000 km. In this calculation, we choose a range of
pathlengths, 6000 - 9000 km, in order to utilize the enhancement associated with
this value.  

 \Tab\ref{table1} gives  our results for SK for an exposure time
of 337.5 Ktyr. We find that, even for the largest allowed value
of $\sin^2 2 \theta_{13}$,  the expectations for NH and IH 
differ by  $2.8 \sigma$ if we have reliable predictions
for neutrino fluxes. If one needs to take ratios to cancel flux 
 uncertainties, then the confidence level in the difference
reduces to $2.2 \sigma$. These low confidence 
levels result from the small event rates
of atmospheric neutrinos at high ({\it i.e} several GeV) energies. Since the observation of the  resonant
amplification of matter effects
requires energies in the $4-10$ GeV range, the 
 larger exposures which are possible at planned
megaton size detectors such as HyperKamiokande (\HK) 
\cite{hk,uno} (higher by a  factor of 6
to 20 compared to SK) are required  to obtain a statistically significant signal. The proposed fiducial volume of \HK~ is 545 Kt.
Hence, with a running time of only 3.3 years, it will be possible to have
a total exposure of 1.8 Mtyr, which is a factor {\bf 6} larger
than what is possible at \SK. In the calculation of the muon 
event rates for \HK, we have used the efficiencies and resolutions of
\SK~\cite{kajita-private1}.

\begin{table}
\begin{center}
\begin{tabular}{| c || c || c c c || c c c |} 
\hline \hline
$\sin^2 2\theta_{13}$ & $\mathrm{N_{vac}}$   
& $\mathrm{N_{NH}}$ 
& $\mathrm{N_{IH}}$ 
& $\mathrm{N_\sigma}$ 
& $\mathrm{R_{NH}}$ 
& $\mathrm{R_{IH}}$ 
& $\mathrm{N_{\sigma_R}}$ \\
\hline\hline
0.05 & 111 
& 101 & 111 & 1 $\sigma$ & 
0.56 & 0.62 & 0.8 $\sigma$ \\
\hline 
0.10 & 110 
& 92 & 107 & {\bf{1.6$\sigma$}} & 
0.52 & 0.6 & {\bf{1.2$\sigma$}} \\    
\hline
0.15 & 108 
& 82 & 103 & 2.3 $\sigma$ & 
0.46 & 0.58 & 2 $\sigma$ \\ 
\hline
0.20 & 108 
& 76 & 100 & 2.8 $\sigma$ & 
0.43 & 0.56 & 2.2 $\sigma$ \\ 
\hline \hline
\end{tabular}
\caption{\footnotesize Integrated muon event numbers for SK (337.5 Ktyr) 
for the energy range E = 5 - 10 GeV and pathlength range 
L = 6000 - 9000 
km. The expected event numbers in the case of  
vacuum oscillations $(\mathrm{N_{vac}})$ and matter oscillations for both  
NH $({\mathrm{N_{NH}}})$ and IH 
$({\mathrm{N_{IH}}})$ are calculated with 
$\sin^2 \theta_{23} = 0.5$, 
$\rm{\vert \da \vert = 0.002~eV^2}$, 
and various different values of $\theta_{13}$.
The number of downward events $\mathrm{N_{Down}} = 178$  
is used in calculating the ratios $\mathrm{R_{NH}} =  
{\mathrm{N_{NH}}}/{\mathrm{N_{Down}}}$ and   
$\mathrm{R_{IH}} =  
{\mathrm{N_{IH}}}/{\mathrm{N_{Down}}}$}
\label{table1}
\end{center}
\end{table}

%
%
\begin{table}
\begin{center}
\begin{tabular}{| c || c || c c c || c c c |} 
\hline \hline
$\sin^2 2\theta_{13}$ & $\mathrm{N_{vac}}$   
& $\mathrm{N_{NH}}$ & $\mathrm{N_{NH}}$ & $\mathrm{N_\sigma}$  
& $\mathrm{R_{NH}}$ & $\mathrm{R_{IH}}$ & $\mathrm{N_{\sigma_R}}$ 
\\ \hline\hline
0.05 & 593
& 537 & 596 & 2.5 $\sigma$  
& 0.56 & 0.63 & 2.3 $\sigma$ \\
\hline 
0.10 & 588 
& 490 & 571 & {\bf{3.7 $\sigma$}}  
& 0.52 & 0.6 & {\bf{2.8 $\sigma$}} \\    
\hline
0.15 & 578  
& 439 & 552 & 5.4 $\sigma$  
& 0.46 & 0.58 & 4.5 $\sigma$ \\ 
\hline
0.20 & 576 
& 406 & 533  & 6.3 $\sigma$  
& 0.43 & 0.56 & 5.1 $\sigma$ \\ 
\hline \hline
\end{tabular}
\caption{\footnotesize Integrated muon event numbers for HK 
(exposure of 1.8 Mtyr) 
for the energy range E = 5 - 10 GeV and pathlength range 
L = 6000 - 9000 km. The expected event numbers in the case of  
vacuum oscillations $(\mathrm{N_{vac}})$ and matter oscillations for both  
NH $({\mathrm{N_{NH}}})$ and IH 
$({\mathrm{N_{IH}}})$ are calculated with 
$\sin^2 \theta_{23} = 0.5$, 
$\rm{\vert \da \vert = 0.002~eV^2}$, 
and various different values of $\theta_{13}$.
The number of downward events $\mathrm{N_{Down}} = 951$  
is used in calculating the ratios $\mathrm{R_{NH}} =  
{\mathrm{N_{NH}}}/{\mathrm{N_{Down}}}$ and   
$\mathrm{R_{IH}} =  
{\mathrm{N_{IH}}}/{\mathrm{N_{Down}}}$}
\label{table2}
\end{center}
\end{table}

\begin{figure}[!h]
\vskip 1cm
{\centerline{
\hspace*{2em}
\epsfxsize=11cm\epsfysize=7.5cm
                     \epsfbox{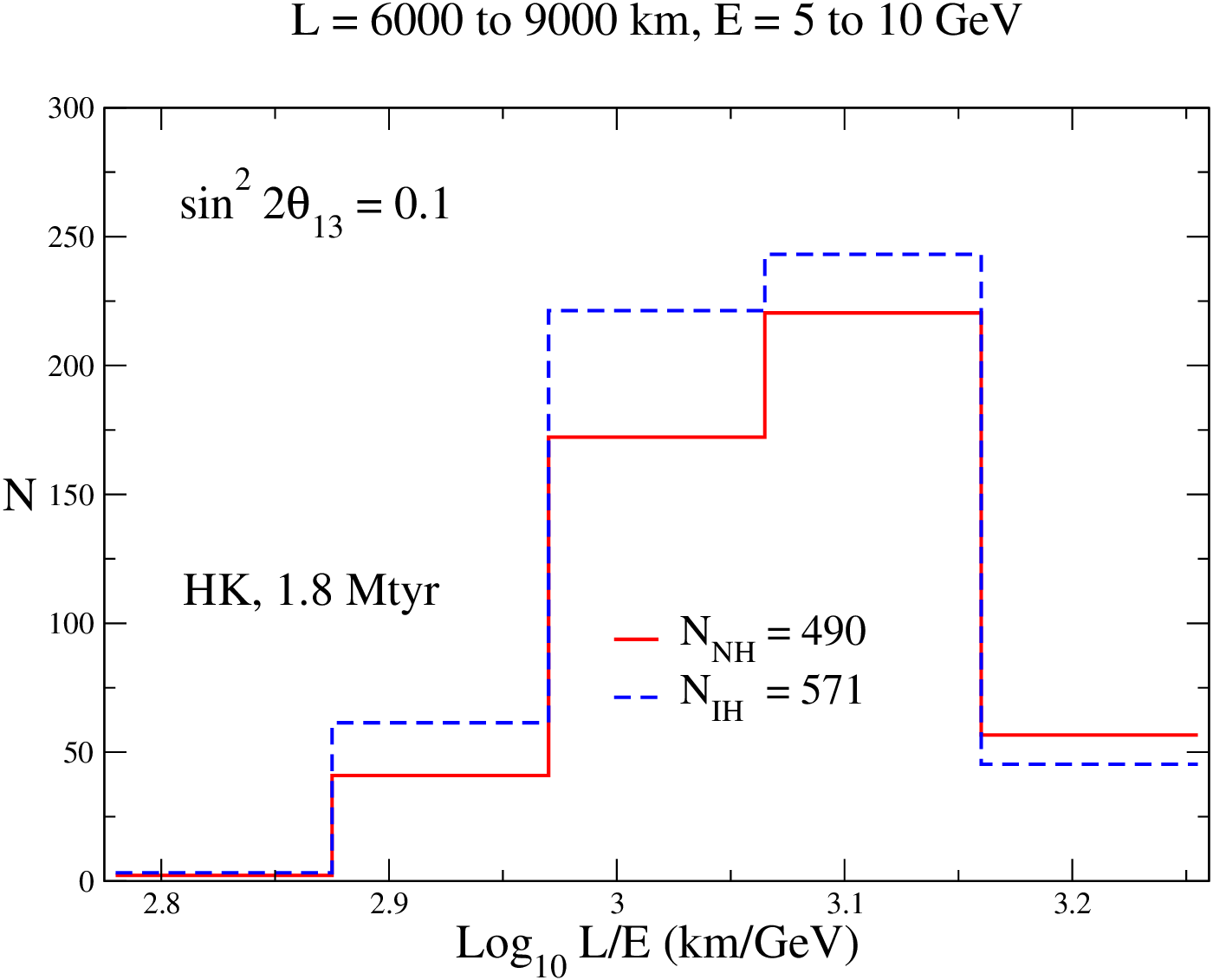}
}
\caption{\footnotesize
Muon ($\sumnumu$) event distribution for \HK(1.8 Mtyr) of matter
oscillations for NH and for IH plotted vs $\rm{Log_{10}(L/E)}$
for the E and L range E = 5-10 GeV,
L = 6000-9000 km.
The values of parameters used are
$\sin^2 \theta_{12} = 0.28$,
$\sin^2 \theta_{23} = 0.5$,
$\stsmall = 0.1$,
$\vert \da \vert = 2 \times 10^{-3}$ \evsq~ and
$\vert \ds \vert = 8.3 \times 10^{-5}$ \evsq. }
\label{fig3}
}
\end{figure}


\fig\ref{fig3} shows the L/E distribution of events for matter oscillations
(for both a NH and an IH) by plotting the events versus $\mathrm{Log_{10}(L/E)}$
for \HK~ in the energy range from 5 - 10 GeV
and pathlength range 6000 - 9000 km,
with $\stsmall = 0.1$. Assuming an L/E resolution similar to 
\SK~ allows a division into five bins in 
$\rm{Log_{10}(L/E)}$ for this L and E range \cite{sk2}.
The drop in $\mathrm{N_{NH}}$ relative to $\mathrm{N_{IH}}$ is 
evident in the third and fourth bins in this figure.
In these energy and zenith angle ranges the  change in $\pmumu$
is predominantly due to the matter effect in $\pmue$.
In \Tab\ref{table2} we present the $\mathrm{N_\sigma}$ values 
calculated using \eq\ref{Nsigma}. We first sum over the events 
of all the 5 bins shown in \fig\ref{fig3} and then compute 
${\mathrm{N_\sigma}}$.
For $\stsmall$ = 0.1 the sensitivity is 3.7$\sigma$.
If instead we consider only the 3rd and 4th bins of
\fig\ref{fig3} and sum the number of events in these two
bins then we get $\mathrm{N_{NH}} = 400$ and $\mathrm{N_{IH}} 
= 475$, which correspond to a sensitivity of 3.8$\sigma$.
Making a narrow choice of bins in (L/E) does not improve
the sensitivity as the effect is distributed over a larger 
L and E range. However,  we will show later that for the L and E ranges 
for which the matter effects in $\pmumu$ arise dominantly 
from $\pmutau$, binning in (L/E) leads to a great improvement
in the sensitivity to mass hierarchy.

As shown in \Tab\ref{table2}, we obtain a $ \gsim 3.7 \sigma$ difference between
$\mathrm{N_{NH}}$ and $\mathrm{N_{IH}}$  
for $\sin^2 2 \theta_{13} \gsim 0.1$, for an exposure of
1.8 Mtyr. Even when we have to take ratios with respect to
downward events, this difference is close to $3 \sigma$.
With an exposure of about 3 Mtyr (which corresponds to 
a running time of about 6 years), one can obtain a 
$3 \sigma$ difference between 
$\mathrm{N_{NH}}$ and $\mathrm{N_{IH}}$, 
for $\sin^2 2 \theta_{13} \simeq 0.05$. 

Next, we check how $\mathrm{N_\sigma}$ and 
$\mathrm{N_{\sigma_R}}$ change with variation
of neutrino parameters. We first illustrate the dependence on
 $\theta_{23}$. For this purpose we fix $\sin^2
2 \theta_{13} = 0.1$ and consider three different values
of $\sin^2 \theta_{23} = 0.4, 0.5$ and $ 0.6$.
The sensitivities for these values of $\theta_{23}$ are
shown in \Tab\ref{table3} for HK exposure of 1.8 Mtyr. The energy
range (5 - 10 GeV) and the pathlength range (6000 - 9000
km) remain the same because $\vert \da \vert$ is 
unchanged. We note from \Tab\ref{table3} that 
the sensitivities are better for $\sin^2 \theta_{23} = 0.6$ 
and worse for $\sin^2 \theta_{23} = 0.4$ compared to the
$\sin^2 \theta_{23} = 0.5$ case. The reason for this is 
 simple. As mentioned earlier, the matter effects in
$\pmumu$ arise mostly due to the matter effects in $\pmue$. 
The matter term in $\pmue$ is proportional to $\sin^2
\theta_{23}$. Therefore, for a larger value of $\sin^2  
\theta_{23}$ we get a larger change in $\pmumu$ due to
matter effects and hence there is a larger difference 
between the expected rates for NH and IH cases.

%
%
\begin{table}
\begin{center}
\begin{tabular}{| c || c || c c c || c c c |} 
\hline \hline
$\sin^2 \theta_{23}$ & $ {\mathrm{N_{vac}}}$ 
& $\mathrm{N_{NH}}$ 
& $\mathrm{N_{IH}}$ 
& $\mathrm{N_\sigma}$ 
& $\mathrm{R_{NH}}$ 
& $\mathrm{R_{IH}}$ 
& $\mathrm{N_{\sigma_R}}$ \\
\hline\hline
0.4 & 599  &
524 & 596  & 3.1 $\sigma$ & 
0.55 & 0.63 & 2.8 $\sigma$ \\
\hline 
0.5 & 588  & 
490 & 571 & {\bf{3.7$\sigma$}} & 
0.52 & 0.6  & {\bf{2.8$\sigma$}} \\    
\hline
0.6 & 596 &
475 & 576  & 4.6 $\sigma$ & 
0.5 & 0.6  & 3.6 $\sigma$ \\ 
\hline \hline
\end{tabular}
\caption{\footnotesize Integrated muon event numbers for HK 
(exposure of 1.8 Mtyr) 
for the energy range E = 5 - 10 GeV and pathlength range 
L = 6000 - 9000 km. The expected event numbers in the case of  
vacuum oscillations $(\mathrm{N_{vac}})$ and matter oscillations for both  
NH $({\mathrm{N_{NH}}})$ and IH 
$({\mathrm{N_{IH}}})$ are calculated with 
$\sin^2 2 \theta_{13} = 0.1$, 
$\rm{\vert \da \vert = 0.002~eV^2}$, 
and various different values of $\theta_{23}$.
The number of downward events $\mathrm{N_{Down}} = 951$  
is used in calculating the ratios $\mathrm{R_{NH}} =  
{\mathrm{N_{NH}}}/{\mathrm{N_{Down}}}$ and   
$\mathrm{R_{IH}} =  
{\mathrm{N_{IH}}}/{\mathrm{N_{Down}}}$}
\label{table3}
\end{center}
\end{table}

We next  study the change in the sensitivities with 
variation in $\vert \da \vert$. We fix $\sin^2 \theta_{23}
= 0.5$, $\sin^2 2 \theta_{13} = 0.1$ and consider three
different values of $\vert \da \vert = 0.001, 0.002$
and $0.003$ \evsq and calculate the muon event
rates for HK exposure of 1.8 Mtyr by integrating over
appropriate energy and pathlength ranges. From  Eq.(\ref{rescon}), we note that 
the resonance occurs at different energies for different
values of $\vert \da \vert$. For $0.003$ \evsq, it occurs 
at $9.5$ GeV. So we choose the energy range of integration
to be $7 - 12$ GeV. For $0.001$ \evsq, the resonance occurs 
at $3.2$ GeV, so we choose the energy range to be $2 - 5$
GeV. But the pathlength range remains unchanged
at 6000 - 9000 km. This is because Eq.(\ref{defL}) 
shows that  the expression for the pathlength for resonant 
amplification of matter effects is independent of 
$\da$. The event numbers and the corresponding sensitivities for this case 
are given in \Tab\ref{table4}. We note that in general one expects a fall the in sensitivity for 
higher values of $\vert \da \vert$ due to the higher resonance energies (and
consequently lower number of total events) \footnote{That this is not the case 
for $\vert \da \vert = 0.002$ compared to $\vert \da \vert= 0.001$ in our
table is attributable to the fact that the energy range for the best fit
value of $\vert \da \vert= 0.002$ has been optimized, while that for the 
other two values has been chosen merely for illustrative purposes.}.

\begin{table}
\begin{center}
\begin{tabular}{| c c c || c c || c c c || c c c |} 
\hline \hline
$\da$ & ${\mathrm{E_{res}}}$ & E Range  
& $\mathrm{N_{Down}}$ & $\mathrm{N_{vac}}$  
& $\mathrm{N_{NH}}$ 
& $\mathrm{N_{IH}}$ 
& $\mathrm{N_\sigma}$
& $\mathrm{R_{NH}}$ 
& $\mathrm{R_{IH}}$ 
& $\mathrm{N_{\sigma_R}}$ \\
\hline\hline
0.001 & 3.2 GeV & 2 - 5 GeV & 
2080 & 1164 & 
1020 & 1125  & 3.3 $\sigma$ & 
0.49 & 0.54 & 2.7 $\sigma$ \\
\hline
0.002 & 6.5 GeV & 5 - 10 GeV & 
951 & 588 & 
490 & 571 & {\bf{ 3.7 $\sigma$}} & 
0.52 & 0.6 & {\bf{ 2.8 $\sigma$}} \\    
\hline
0.003 & 9.5 GeV & 7 - 12 GeV &
598 & 434 &
356 & 398 &  2.2 $\sigma$ & 
0.59 & 0.67 & 1.8$\sigma$ \\ 
\hline \hline
\end{tabular}
\caption{\footnotesize Muon event numbers for HK for an
exposure of 1.8 Mtyr for different values of $\vert \da \vert$.
The pathlength range, L = 6000 - 9000 km, is the same for all
three cases but the energy ranges are different because the
resonance energies are different. The expected event numbers 
in the case of vacuum oscillations $(\mathrm{N_{vac}})$ and
 matter oscillations for both  
NH $({\mathrm{N_{NH}}})$ and IH 
$({\mathrm{N_{IH}}})$ are calculated with 
$\sin^2 \theta_{23} = 0.5$ and $\sin^2 2 \theta_{13} = 0.1$ 
Because of different energy ranges, the number of downward 
events in each case is different and are listed in table.
They are used in calculating the ratios $\mathrm{R_{NH}} =  
{\mathrm{N_{NH}}}/{\mathrm{N_{Down}}}$ and   
$\mathrm{R_{IH}} =  
{\mathrm{N_{IH}}}/{\mathrm{N_{Down}}}$}
\label{table4}
\end{center}
\end{table}

\begin{figure}[!h]
\vskip 1cm
{\centerline{
\hspace*{2em}
\epsfxsize=11cm\epsfysize=7.5cm
                     \epsfbox{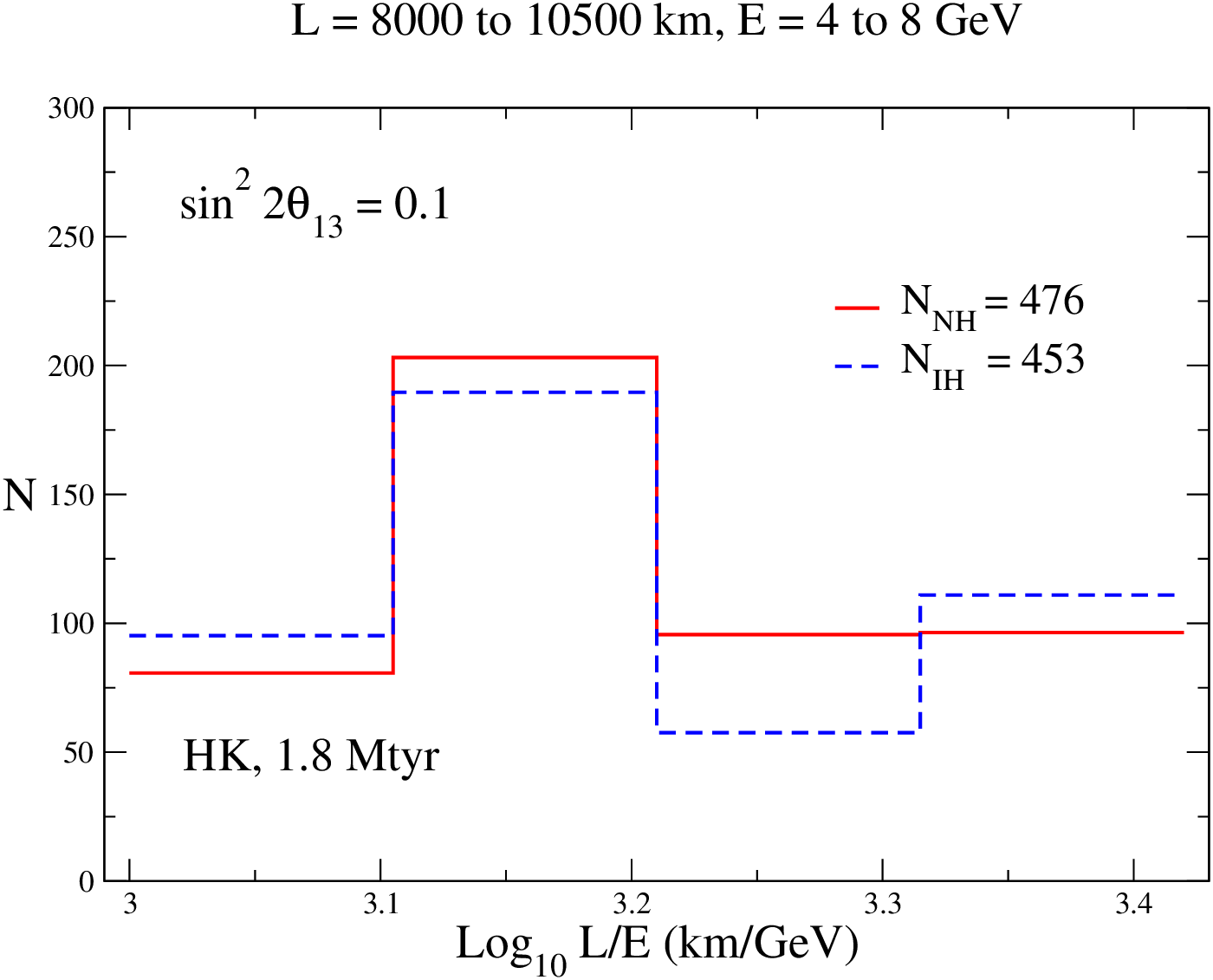}
}
\caption{\footnotesize 
Muon ($\sumnumu$) event distribution for \HK(1.8 Mtyr) in matter 
and in vacuum plotted vs $\rm{Log_{10}(L/E)}$ 
for the 
E and L range E = 4-8 GeV, L = 8000-10500 km. 
The values of parameters used are 
$\sin^2 \theta_{12} = 0.28$, 
$\sin^2 \theta_{23} = 0.5$, 
$\stsmall = 0.1$,  
$\vert \da \vert = 2 \times 10^{-3}$ \evsq~ and 
$\vert \ds \vert = 8.3 \times 10^{-5}$ \evsq. }
\label{fig4}
}
\end{figure}

It was shown in references \cite{us1,us2} that for certain
ranges of L and E, $\pmumu$ is larger compared to its 
vacuum value, due to a sharp fall (by as 
much as $70 \%$) in $\pmutau$. For $\vert \da \vert = 0.002$
\evsq, this occurs in the energy range 4 - 6 GeV and pathlength
range 8000 - 10500 km. In \fig\ref{fig4} the total muon event rates
are computed and plotted as function of L/E for values of
$\mathrm{Log_{10}(L/E)}$ in the range 3.0 - 3.4. If we assume 
\SK~ resolutions, this range can be divided into four bins.
We see that for the bin with $\mathrm{Log_{10}(L/E)} = 3.2 - 3.3$
the change due to the sign of $\da$ is particularly pronounced.
Note also that, as mentioned above, unlike the previous ranges, here {\it the 
expected number of muon events for the case of NH is {\bf larger}
than that for the case of IH}.
This phenomenon, however, occurs only for narrow ranges in 
energy and pathlength and hence one needs a detector with 
good L/E resolution to observe it.  
Given a detector with such resolution however, the difference in the expected muon rates in the bin
with $\mathrm{Log_{10}(L/E)} = 3.2 - 3.3$ is an excellent 
indicator of the
neutrino mass hierarchy. From the event numbers given in 
\Tab\ref{table5}, we see that the difference in this bin
corresponds to a ${\bf 4 \sigma}$ signal for $\stsmall = 0.1$ with an 
exposure of just 1.8 Mtyr in HK. If we sum the events in
the two bins with $\mathrm{Log_{10}(L/E)} = 3.1 - 3.2 \ 
{\rm and} \ 3.2 - 3.3$, the difference between the expectations 
for NH and IH corresponds to a sensitivity of 3$\sigma$. 

 
%
\begin{table}
\begin{center}
\begin{tabular}{| c || c || c c c || c c c |} 
\hline \hline
$\sin^2 2\theta_{13}$ & $\mathrm{N_{vac}}$   
& $\mathrm{N_{NH}}$ & $\mathrm{N_{IH}}$ & $\mathrm{N_\sigma}$  
& $\mathrm{R_{NH}}$ & $\mathrm{R_{IH}}$ & $\mathrm{N_{\sigma_R}}$ 
\\ \hline\hline
0.05 & 40 
& 68 & 46 & 2.6 $\sigma$  
& 0.21 & 0.14 & 2.5 $\sigma$ \\
\hline 
0.10 & 41 
& 96 & 58 & {\bf{3.9 $\sigma$}}  
& 0.3 & 0.18 & {\bf{3.4$\sigma$}} \\    
\hline
0.15 & 49 
& 130 & 74 & 4.9 $\sigma$  
& 0.4 & 0.23 & 4.1 $\sigma$ \\ 
\hline
0.20 & 50 
& 158 & 86  & 5.7 $\sigma$  
& 0.49 & 0.26 & 4.8 $\sigma$ \\ 
\hline \hline
\end{tabular}
\caption{\footnotesize Integrated muon event numbers for HK 
(exposure of 1.8 Mtyr). The events selected are those in the   
energy range E = 4 - 8 GeV and pathlength range 
L = 8000 - 10500 km, with $\mathrm{Log_{10}(L/E)} = 
3.2 - 3.3$. The expected event numbers in the case of  
vacuum oscillations $(\mathrm{N_{vac}})$ and matter oscillations for both  
NH $({\mathrm{N_{NH}}})$ and IH 
$({\mathrm{N_{IH}}})$ are calculated with 
$\sin^2 \theta_{23} = 0.5$, 
$\rm{\vert \da \vert = 0.002~eV^2}$, 
and various different values of $\theta_{13}$.
The number of downward events $\mathrm{N_{Down}} = 325$  
is used in calculating the ratios $\mathrm{R_{NH}} =  
{\mathrm{N_{NH}}}/{\mathrm{N_{Down}}}$ and   
$\mathrm{R_{IH}} =  
{\mathrm{N_{IH}}}/{\mathrm{N_{Down}}}$}
\label{table5}
\end{center}
\end{table}


\Tab{\ref{table5}} also gives the number of events in matter ($\da>0$)
and the values of $\sigma$
sensitivity for this $\mathrm{Log_{10}(L/E)}$ bin for three values of
$\theta_{13}$. This shows a higher sensitivity as compared to 
\Tab\ref{table2} and more than 2.5$\sigma$ sensitivity can 
be achieved for $\stsmall = 0.05$ with the exposure considered here.

\section{Conclusions}

\begin{itemize}
\item
The neutrino mass hierarchy, presently unknown, is a powerful discriminator
 among various classes of unification theories. 
We have shown that the total muon ($\sumnumu$) survival rate in
atmospheric events can provide a novel and useful method of determining the
hierarchy in megaton water \chr detectors\footnote{Even though we have
 presented the results for \HK, our conclusions should remain valid for
 similar detectors such as \UNO~ \cite{uno} or 
the one proposed in Frejus tunnel \cite{frejus} in Europe.}. 
\item 
The rates for both normal and inverted hierarchies  differ from the vacuum
 rate and from each other due to large matter sensitive variations 
not only in $\pmue$ but also in $\pmutau$. 
We have identified energy and pathlength ranges where these effects can 
be fruitfully observed in a statistically significant manner for 
running times of $\sim 3-4$ years for  values of $\sin^2\theta_{13}\geq 0.05$.
\item
We have calculated the $\sigma$ sensitivities for normal and inverted
hierarchies via differences in absolute events  as well as differences in event
ratios (to eliminate  uncertainties in the atmospheric flux) for various
choices of the neutrino parameters. We find that the energy ranges $5-10$ Gev
and $4-8$ GeV, when combined with the length ranges $6000-9000$ km and
$8000-10500$ km respectively{\footnote{In the latter case, as discussed in the context of Figure {\ref{fig4}},
the effect is apparent in the L/E bin specified by $\mathrm{Log_{10}(L/E)}=3.2-3.3$.}}, 
allow for a statistically significant
determination of the hierarchy. The first (L,E) range combination allows
us to  probe the $\pmue$-induced matter 
effects in $\pmumu$, while the second combination provides a way of detecting
the $\pmutau$ effects in $\pmumu$.
\item
Finally, we have also studied the dependence  of our results on the
uncertainties in the presently known values of $\sin^2 \theta_{23}$ and  
$\rm{\vert \da \vert}$. 
\end{itemize}

\vskip 10pt
\section*{Acknowledgements}
The authors thank T.~Kajita for useful communications and S.~Choubey
for useful discussions regarding the statistical sensitivity for Normal and Inverted hierarchy. 
P.M. wants to thank the Weizmann Institute of Science, Israel for 
hospitality. RG would like to thank the Stanford Linear Accelerator Center 
(SLAC) and the Institute for Nuclear Theory (INT) at the University of
Washington for hospitality while this work was in progress. 


\end{document}